 \documentstyle[aps,epsfig,tighten]{revtex}
\begin{document}
 
 \title{Intrinsic limitations on the size of quantum
 database\thanks{email address: gllong@mail.tsinghua.edu.cn}}
 \author{Gui Lu Long$^{1,2,3}$, Yan Song Li$^1$, Wei Lin Zhang$^1$,
 Chang Cun Tu$^1$ }
 \address{$^1$Department of Physics, Tsinghua University, Beijing, 100084, P.
 R. China\\
$^2$Institute of Theoretical Physics, Chinese Academy of Sciences,
Beijing, 100080, P. R. China\\
$^3$Center of Nuclear Theory, National Laboratory of Heavy Ion Physics,
Lanzhou 730000, P. R. China}
 \date{\today \vspace*{-.3cm}}
 \maketitle

%%%%%%%%%%%%%%%%%%%%%%%%%%%%%%%%%%%%%%%%%%%%%%%
 \def\mywidth{\global\columnwidth25.5pc
          \global\hsize\columnwidth\global\linewidth\columnwidth
          \global\displaywidth\columnwidth}
 
 \def\be{\begin{equation}}
 \def\ee{\end{equation}}
 \def\bea{\begin{eqnarray}}
 \def\eea{\end{eqnarray}}
 \def\Adag{A^\dagger}
 \def\g{\gamma}
 \newcommand{\mattwoc}[4]{\left[
          \begin{array}{cc}{#1}&{#2}\\{#3}&{#4}\end{array}\right]}
 \newcommand{\ket}[1]{\mbox{$|#1\rangle$}}
 \newcommand{\bra}[1]{\mbox{$\langle #1|$}}
 \def\>{\rangle}
 \def\<{\langle}
 \newcommand{\mypsfig}[1]{\psfig{file=#1}}
 
 \begin{abstract}
 It is found that Grover's quantum search algorithm is not robust
against phase inversion and Hadmard transformation inaccuracies.
  Imperfect phase inversions and Hadmard-Walsh transformations 
in Grover's quantum search algorithm 
  lead to reductions in the maximum probability of the marked
  state and affect the efficiency of the algorithm. even in the
  absence of decoherence. Given the degrees of inaccuracies, we
  find that to guarantee half rate of success, the size of the database
  should be in the order of $O({1 \over \delta^2})$, where $\delta$ is the
  uncertainty.
 \end{abstract}
 
 %\begin{multicols}{2}
 
 \section{Introduction}
 Grover's quantum search algorithm is a remarkable achievement in
 quantum computing\cite{r1}. There have been intensive interests in Grover's 
quantum search algorithm recently
\cite{r2,r3,r4,r5,r6,r7,r8,r9,r10,r11,r12,r13,r14,r15,r16,r17,r18}
It uses only two simple gate operations, the
 controlled phase rotations and Hadamard transformations.
 It has been successfully demonstrated in
 solution NMR bulk quantum computers with a few qubits \cite{r2,r3}.
 However, the inevitable quantum state decoherence and gate inaccuracies can
 introduce errors\cite{r18,r19}, 
which accumulate through the computation and make
 long computation unreliable.
 While, in order to find out the marked state with high probability, it
 still requires exponential number of iterations. Then, the error
 probability of the complete algorithm may be as exponentially large as
 the error probability of each iteration. In other words, even with
 small imperfection per step, large scale quantum search may be
 difficult.
 
 Fortunately, recent study in quantum error correction shows that in
 principle, whenever the noise rates are below a constant threshold, an
 arbitrary long quantum operations can be performed reliably through
 {\em fault-tolerant quantum computation} \cite{r20}.
 Experimentally, different types of faults can occur with different
 rates and will affect the efficiencies of the algorithm differently. 
For example, 
 the effect of quantum state decoherence and operational errors on the
 efficiency of quantum algorithms have been studies in
 \cite{r21} with ion trap quantum computers. A
 good understanding of the effect from different noises on the
 algorithms can help us look for specific potential physical
 realizations of quantum computers. 

 In this paper, we address the problem of
 influences of imperfect gate operations in the quantum search
 algorithm, in the absence of decoherence and error corrections.
 We will show that systematic 
phase mismatching 
and random errors in the Walsh-Hadmard transformation
lead to exponential reduction in the
 maximum success probability when $n$ is linearly increased.
 Therefore, if we can not avoid them completely, to ensure
 a large success rate in a quantum searching machine, the size of the
 database should be limited.
 This limitation is due to the
 intrinsic vulnerability of the algorithm to imperfect gate
 operations. In designing a quantum searching machine, this limitation
 should be taken into account.

 The paper is organized as follows: Section 2 is devoted to the
description of
 different error models in phase mismatching and the corresponding
 simulation results.  In section 3, we present the consequence of
 imperfect Hadamard transformation. Section 4 gives a short summary.

 \section{effects of imperfect phase inversions}
 Grover's algorithm consists of essentially four steps in an
 iteration\cite{r5}: (1) a Walsh-Hadamard transformation $U=W$; (2) a
 phase inversion of the prepared state $|\gamma\rangle$, usually
 $|\gamma\rangle=|0\rangle$, $I_{\gamma}=I-2|\gamma\rangle\langle
 \gamma|$; (3) a phase inversion of the marked state $|\tau\rangle$,
 $I_{\tau}=I-2|\tau\rangle\langle\tau|$ ; and (4) an inverse of the
 Walsh-Hadamard transformation $U^{-1}=W$ ($W$ is self-inverse.). The
 operator for one Grover iteration is $Q=-I_\gamma U^{-1}I_\tau U$.

 In this section, we focus on the imperfection in
 phase inversions and therefore choose $U$ to be the ideal Hadamard
 transformation.  We consider the imperfections in the phase
 inversion to be {\em systematic}, such that
 \begin{eqnarray}
  I_\gamma& =&I-(1- e^{i\theta }) |\gamma \rangle \langle \gamma |,\nonumber\\
  I_\tau& =&I-(1-e^{i\varphi }) |\tau \rangle \langle \tau |,\nonumber\\
 \end{eqnarray}
 where $\theta = \pi + \theta_0, \varphi = \pi+\varphi_0$ with
 $\theta_0$ and $\varphi_0$ {\em constant} and small.  When
 $\theta_0=\varphi_0=0$, we recover the original Grover's algorithm.
 The generalized quantum search algorithm is a rotation in a
 2-dimensional space spanned by $|\gamma\rangle$ and
 $|\tau\rangle$. In the following two orthonormal basis
 \begin{eqnarray}
   |1\rangle &=&(|\gamma\rangle-U_{\tau \gamma }
   U^{-1}|\tau \rangle)\over\sqrt{1-|U_{\tau\gamma}|^2},\nonumber\\
   |2\rangle &=&U^{-1}|\tau \rangle .
 \label{eb}
 \end{eqnarray}
 with $U_{\tau\gamma} = \<\tau|U|\gamma\> = 1/\sqrt{N}$,
 the operator $Q$ is represented by
 \begin{eqnarray}
  \left(\begin{array}{cc}
  -e^{i\theta }-| U_{\tau \gamma }| ^2( 1-e^{i\theta })
  &(1-e^{i\theta }) U_{\tau \gamma}\sqrt{1-|U_{\tau\gamma}| ^2}\\
  e^{i\varphi }( 1-e^{i\theta })U_{\tau \gamma }^*\sqrt{1-|U_{\tau\gamma }
  | ^2}&-e^{i\varphi}[1-(1-e^{i\theta})| U_{\tau \gamma }|^2]
  \end{array}
  \right)
 \end{eqnarray}

 Let $\delta = \theta - \varphi = \theta_0 - \varphi_0$. 
It has been shown that to construct an efficient quantum search
algorithm, $\theta$ and $\varphi$ must equal to one
another\cite{r22,r23}. However due to imperfections in gate operations,
this phase matching requirement can not be strictly satisfied. In the
 following, we show that nonzero constant $\delta$ results in
 exponential reduction in the maximum success probability of Grover's
 algorithm asymptotically.
 
 Since both $\theta_0$ and $\varphi_0$ are small, dropping
 off an overall phase, we approximate $Q$ as
 \be
         Q \doteq
         \cos{\delta} I + i \sin{\delta} \sigma_z + i\beta' \sigma_y +
         o(\beta')
 \,,
 \ee
 where $\sigma_x, \sigma_y$ and $\sigma_z$ are
 Pauli operators and $I$ is the identity operator in dimension 2.
 $\beta' = 2\beta + O(\theta_0 \beta) = 2\beta + o(\beta)$ with $\beta
 = \frac{\sqrt{N-1}}{N}$.
 For small $\delta$ , we can further
 simplify operator $Q$ as,
 \begin{eqnarray}
  Q\approx I+iG\approx
  e^{iG},
 \nonumber
 \end{eqnarray}
 with $G = \sin{\delta} \sigma_z + \beta'\sigma_y$.
 Using $G^2=(\delta^2+\beta^{\prime2})I$, we obtain
 \be
  Q^j = \left[ \begin{array}{cc}
  \cos{j \lambda} + i\frac{\delta \sin{j\lambda}}{\lambda} &
  \frac{\beta' \sin{j\lambda}}{\lambda} \\
  -\frac{\beta' \sin{j\lambda}}{\lambda} &
  \cos{j \lambda} - i\frac{\delta \sin{j\lambda}}{\lambda}
  \end{array} \right]
 \,,
 \label{e16}
 \ee
 with $\lambda = \sqrt{\delta^2 + \beta'^2}$.
 Then, starting from the prepared state
 $|\gamma\rangle=
 \sqrt{1-|U_{\tau\gamma}|^2}|1\rangle+U_{\tau\gamma}|2\rangle=\cos\beta
 |1\rangle +\sin\beta |2\rangle\approx |1\rangle$, after $j$ number of
 iterations, the norm of the amplitude of the marked state in the
 quantum computer is
 \begin{eqnarray}
  |B_j|\approx {\beta' \over \lambda}\sin(j\lambda).
 \label{e17}
 \end{eqnarray}
 and the maximum probability of the marked state in the algorithm is
 %
 %reduced to
 %As in Grover's initial algorithm, with optimal iteration steps,
 % $\sin(j\lambda)\approx 1$. However, the maximum probability of the
 % marked state in the algorithm is reduced to,
 %
 \begin{eqnarray}
 \label{e3}
  P_{max}\approx {\beta^{\prime2} \over\ {\beta^{\prime2}+\delta^2} }\le 1
 \,.
 \end{eqnarray}
 Therefore, for large $N$, Grover's algorithm is efficient only
 when
 $\delta=0$. When $\delta\neq 0$, we find
 \be
          P_{max} \approx \frac{\beta'^2}{\delta^2} \sim \frac{4}{N\delta^2}
 \,.
 \ee
 Thus, $P_{max}$ decreases linearly with $N$ or exponentially with
 $n=\log_2{N}$.  This concludes our proof that systematic phase
 mismatching results in exponential reduction in the success
 probability and consequently gives an upper bound on the size of the
 database. If half rate of success is required, that is $P_{max}\ge
 1/2$, $N$ cannot exceed $8/\delta^2$.

 So far, we have assumed that the errors in the phase
 inversions are systematic such that $\delta$ is constant. We
 now extend this simple model (EM1) to another two
 error models.
 The second error model (EM2) assumes $\delta$ in each step is a
 Gaussian random variable with mean $\delta_0 = 0$ and standard
 deviation $s$. Such an error is conventionally defined as {\em random} error.
 Finally, we let $\delta$ be a Gaussian random variable with mean
 $\delta_0 \neq 0$ and standard deviation $s$ (EM3).
 The exact effect of EM2 and EM3 are difficult to compute analytically
 due to their randomness.  Hence, we only present the simulation
 results.
 We vary $n=Log_2N$ and run the algorithm with sufficient number of
 iterations so that a maximum probability is found.
 Since $\delta$ in EM2 and EM3 are random variables, we adopt the {\em
 random sampling} techniques in the
 simulation.  The relationships between the maximum success
 probabilities and the size of the database are shown in Fig. 2 and
 Fig. 3 for EM2 and EM3 respectively.
 For comparisons, we also provide the simulation result from
 EM1 in Fig.1.

 Our simulation results are consistent with mathematical
 predictions. First, both systematic and random errors cause reduction
 in the maximum probability.
 Second, the success probability drops quickly after a transition
 point, which is determined by the error parameter $\delta_0$ and $s$.
 When $n$ is large, the probabilities decreases exponentially.
 Third, the different effects of systematic errors and random errors also
 meet our expectations. Mathematically, systematic errors cause the
 error amplitudes to grow exponentially with the number of gates
 applied; while the random errors cause the error probabilities
 to grow linearly
 This difference is clearly demonstrated in our simulation
 results.
 Fig. 2 shows that random errors give a much larger transition point
 than systematic errors.
 Fig. 3 shows that the average
 success probability from EM3 is nearly identical to EM1 except some small
 fluctuations.

It is shown in this section that systematic errors in the phase
inversions lead to reduction in the maximum probability of finding the
marked state. Random erros also affect this successfl rate, but in a
lesser degree.
 In practice, we should make $\delta_0$ as small as possible.  However,
 due to imperfection, nonzero $\delta_0$ occur inevitably.  due to
 imperfection, nonzero $\delta$ occur inevitably.  For instance,
 systematic errors arises from imperfect calibration and inhomogeneity
 in the radio frequency pulses in NMR realization.  Random errors are
 always present in a realistic environment.  These errors will reduce
 the maximum probability of the algorithm.
 To make an estimate on the
 combined effect of systematic and random errors(EM3), we assume that
 random errors affect the algorithm just like the systematic errors.
 Then we can treat $\Delta=2\delta$ as the uncertainty due to both
 systematic errors and random errors and use this to derive an upper
 bound for the size of a quantum database:{\it any phase inversion
 operation is imperfect, there is an uncertainty, and this uncertainty
 sets an upper bound on the size of the database $N$. }  For half rate
 success, the dimension of the database should be less than ${64 \over
 \Delta^2}$.
 
 \section{Imperfect Hadamard transformation}
Hadmard-Walsh transformations are also subject to errors. To study the
effect of the imperfect Hadmard-Walsh transformation, 
 let's take $\delta=0$ in eqn. (\ref{e16}).  Then the maximum
 probability for finding the marked state is approximately
 $\sin^2(2j\beta)$ for perfect unitary transformation.
For perfect Hadamard-Walsh transformation,
 $\beta=\arcsin(|U_{\tau\gamma}|)$, and $|U_{\tau\gamma}|=\sqrt{1/N}$.
For systematic errors in the Hadmard-Walsh transformation,
the matrix elements of $U$ is no
longer equal to $\sqrt{1/N}$.
If 
 $|U_{\tau\gamma}|$ is larger than $\sqrt{1\over N}$, then the
 algorithm will require less steps in reaching the desired state
 compared with the standard Grover's algorithm.  If it is smaller than
 $\sqrt{1\over N}$, the algorithm will require more steps of iteration.
 In this case, the searching algorithm
still can give a probability quite close to unity.
But if one makes a measurement at the normal optimal number of
iteration, one will gets a reduction in the successful rate. This
difficulty can be overcome by using the algorithm several times with
measurements made around the optimal iteration which is similar to the
one used in Ref.\cite{r12}.

Here, we can give a simple interpretation why Grover's algorithm is
optimal. The rigorous proof has been given in Refs.\cite{r11}.
 Grover's algorithm can be seen as a rotation of the state vector in a
 2-dimensional space span by $|\tau\rangle$ and $\gamma\rangle$. Each
 iteration rotates an angle
 $\lambda=\beta'=2\sin(\theta/2)\beta$. $\theta=\phi=\pi$ gives the
 largest angle $2\beta=2\arcsin(|U_{\tau\gamma}|$. So one has to choose
 phase inversions.  As for the unitary transformation $U$, at first
 glance one maybe attempted to think that a larger $|U_{\tau\gamma}|$
 will constitute a faster search algorithm.
 However, since $U$ is unitary, its matrix elements satisfy the
 normalization relation $\sum_{\tau}|U_{\tau\gamma}|^2=1$, where $\tau$
 runs through all the $N$ basis states. The mean value of the matrix
 element is $\sqrt{1\over N}$.  If some of the matrix elements are
 larger than the average, some other matrix elements will be less than
 this average.  In other words, while making the search for some marked
 states in less steps, the modified algorithm has to search the rest of
 the basis states in more steps.  In contrast, the original Grover
 algorithm searches all possible marked state with the same optimal
 number of iterations.  Together with its simpleness and easy
 implementation, the Walsh-Hadamard transformation lend itself the best
 choice.
 
 We discuss the effects of random errors in the Walsh-Hadamard 
transformation in a simple model. In this case, the algorithm is no
longer a simple rotation in 2 dimensions. Though in each iteration, the
operator can be approximately written as
\begin{eqnarray}
Q=\left(\begin{array}{cc} \cos\beta & \sin\beta \\
                    -\sin\beta & \cos\beta \end{array}\right),
\end{eqnarray}
the basis states in each iteration has been changed, that is, the 2
dimensional space in each iteration is no longer the same. This is
apparent by inspecting the expressions in eqn. (\ref{eb}). Suppose in
the first iteration, the unitary transformation is $U$ and in the
following iteration,  the operator becomes $V$. Then after the first
iteration, the state vector of the quantum computer is
\begin{eqnarray}
|\psi_1\rangle =\cos\beta |1\rangle-\sin\beta|2\rangle\approx \cos\beta
|1'\rangle -\sin\beta U^{-1}V|2'\rangle,
\end{eqnarray}
where $|2'\rangle =V^{-1}|\tau\rangle$. Because $U\neq V$, $U^{-1}V$ is
no longer the identity operator. Expanding
$U^{-1}V|2'\rangle=(U^{-1}V)_{22}|2'\rangle +$ ..., 
we see that the Grover search operator acts only on the subspace span by
$|1'\rangle$ and $|2'\rangle$, and the other terms are leaked out the 2
dimensional space. To make an estimate, let's assume that in each
iteration,$(U^{-1}V)_{22}|2'\rangle\approx (1-\delta_1)|2\rangle$ +
higher order terms. Then in this model, the matrix for a Grover search
operator becomes
\begin{eqnarray}
Q=\left(\begin{array}{cc} \cos\beta & \sin\beta(1-\delta_1) \\
                    -\sin\beta & \cos\beta (1-\delta_1)\end{array}\right).
\end{eqnarray}
Starting from initial state $|\gamma\rangle\approx |1\rangle$, after $j$
iterations, the amplitude of the state $|2\rangle$ becomes
\begin{eqnarray}
\left| (1-{j-1 \over 2}\delta_1) \sin(j\beta)\right|,
\end{eqnarray}
where only first order in $\delta_1$ is retained. With optimal number of
iterations,  $j\approx{\pi\sqrt{N} \over 4}$, $\sin(j\beta)\approx 1$,
the successful rate is
\begin{eqnarray}
P\approx \left(1-{\pi\sqrt{N}\delta_1 \over 8}\right)^2\approx
1-{\pi\sqrt{N}\delta_1 \over 4}.
\end{eqnarray}
For half success rate, one must have $N \le {4 \over \pi^2\delta_1^2}$,
which is similar to the limitation on the size of the database in the
phase inversion inaccuracies. However, the mechanism is different.
Here the random errors play a more
important role than the systematic errors, whereas in the phase
inversion case, it is just the opposite.
 
 \section{Summary}
 In summary, we find that the dominating gate imperfection in Grover's
 algorithm is the systematic phase mismatching and the random errors in
the Walsh-Hadmard transformation. 
Using the results obtained in this work, it is easy to understand the
simulating 
results of Ref.\cite{r21}. In Fig.1a of \cite{r21} is the results with
only random errors(both in phase inversions and Hadmard transformation),
we see that the peak in the probability curve drops down
as random errors grow. But the position of the peak is relatively fixed.
Random errors in the phase inversions does not affect the algorithm very
seriously. Random errors in the Hadmard transformation reduce the
maximum probability. The optimal iteration number remain more or
less the same.
When there is only systematic errors as shown in Fig.2b of Ref.\cite{r21},
we see  a drop in the
maximum probablity and also a shifting of the peak position to the left,
as is shown in Fig.2b of Ref.\cite{r21}.
The drop in maximum probability is caused by phase mismatching. The
shift of the peak position is due to the systematic errors in the
Walsh-Hadmard transformation. 

These gate inaccuracies set an upper bound on the
 size of the database. We estimate that the upper bound is inversely
 proportional to the quadrature of the uncertainty in the phase
 mismatching or in the Walsh-Hadmard transformation.
 In real quantum computation, imperfect gate operations exist all the
 time at constant rate while decoherence increases rapidly with
 computing time.  
 At the early stage of a quantum computation, gate
 imperfection is dominant in affecting a quantum algorithm. As the
 computation continues, decoherence increases and then dominates.
 Suitable quantum correction codes and in particular fault-tolerant
quantum computation can reduce the decoherence, and ease the stringent
requirement on gate accuracies.
The limitations on the quantum datasize can then be greatly relieved.
 
 Encouragements from Prof. Haoming Chen and Hongzhou Sun are
 gratefully acknowledged.

\noindent Figure captions
%\begin{figure}
%\begin{center}
%\epsfig{figure=S1.EPS,width=5cm}
%\end{center}
\noindent {Fig.1. EM1 with $\delta_0 = 10^{-2},10^{-3},10^{-4}$.}
%\end{figure}

%\begin{figure}
%\begin{center}
%\epsfig{figure=S2.EPS,width=10cm}
%\end{center}
\noindent{Fig.2. EM2 with $\delta_0 = 0, s = 10^{-2}$ and $\lambda =\beta'$}
%\end{figure}

%\begin{figure}
%\begin{center}
%\epsfig{figure=S3.EPS,width=10cm}
%\end{center}
\noindent{ Fig.3. EM3 with $\delta_0 = 10^{-2},10^{-3},10^{-4},
s=10^{-3}$}
%\end{figure} 
 

\begin{thebibliography}{99}
 \bibitem{r1} L. K. Grover,{\it Quantum mechanics helps in searching for  a needle in a haystack}, Phys. Rev. Lett. 79 (1997) 325.
 \bibitem{r2} J.A. Jones, M. Mosca, R. H. Hansen,  {\it Implementation of a quantum search algorithm on a quantum  computer},   Nature, 393 (1998) 344.
 \bibitem{r3} I.L. Chuang, N. Gershenfeld, M. Kubinec,  {\it Experimental implementation of fast quantum searching},  Phys. Rev. Lett.  80 (1998) 3408.
\bibitem{r4} L. K. Grover, {\it Quantum computers can search arbitraryily large databases by a single query}, Phys. Rev. Lett., 79 (1997) 4709.
\bibitem{r5} Lov K. Grover, {\it Quantum computers can search rapidly by  using almost any transformation}, Phys. Rev. Lett. 80 (1998) 4329.
\bibitem{r6}  G. Brassard, {\it Searching a quantum phone book}, Science, 275 (1997) 627.
\bibitem{r7} G. Brassard, P. H\o yer and A. Tapp, {\it Quantum counting} Lanl-eprint/quant-ph/9805082. 
\bibitem{r8} Liping Fu, Li Xiao, Jun Luo and Xizhi Zheng,to appear in Chin. Phys. Lett..
\bibitem{r9} C. Bennett et al, {\it Strengths and weaknesses of quantum computing}, Lanl-eprint/quant-ph/9701001, also in SIAM journal on Computing.
\bibitem{r10} M. Boyer, G. Brassard, P. H\o yer, A. Tapp,{\it Tight bounds on quantum searching}, Lanl-eprint/quant-ph9605034; also in Fortsch. Phys. 46 (1998) 493.
\bibitem{r11} C. Zalka, {\it  Grover's quantum searching algorithm is optimal}, Lanl-eprint / quant-ph /9711070.
\bibitem{r12} C. Zalka, {A Grover-based quantum search of optimal order for an umknown number of marked states}, Lanl-eprint/quant-ph/9902049.
\bibitem{r13} D. Biron et al, {\it Generalized Grover search algorithm for arbitrary initial amplitude distribution},  Lanl-eprint/quant-ph/9801066.
\bibitem{r14} A. Kumar Pati, {\it Fast quantum search algorithm and bounds on it}, Lanl-eprint/quant-ph/9807067.
\bibitem{r15} Y. Ozhigov, {\it Speedup of iterated quantum search by parallel performance}, Lanl-eprint/quant-ph/9904039. 
\bibitem{r16} R. Gingrich, {\it Generalized quantum search with parallelism},C. P. Williams and N. Cerf,Lanl-eprint/quant-ph/9904049. 
\bibitem{r17} R. Josza, {\it  Searching in Grover's algorithm}, Lanl-eprint/quant-ph/9901021.
\bibitem{r18} Sixia Yu and Chang-pu Sun, {Quantum searching's underlying SU(2) structure and its quantum decoherence effects}, Lanl-eprint/quant-ph/9903075.
 \bibitem{r19} I.L. Chuang, R. Laflamme, P. W. Shor and W.H. Zurek, {\it  Quantum computers, factoring and decoherence}, Science, 270 (1995) 1633.
 \bibitem{r20} J. Preskill, {\it Reliable quantum computers}, Proc. R. Soc. London A454 (1998) 385.
 \bibitem{r21} K. Obenland and A. M. Despain, {\it Simulating the effect of decoherence and inaccuracies on a quantum computer},quant-ph/9804038.
\bibitem{r22} G.L. Long, W.L. Zhang, Y. S. Li and L. Niu, {\it Arbitrary phase rotation of the marked state cannot be used in Grover's quantum search algorithm},
Lanl-eprint/quant-ph/9904077, also in Commun.Theor. Phys. 32 (1999) 335.  
\bibitem{r23} G.L. Long, Y. S. Li, W.L. Zhang and L. Niu, {\it Phase matching in quantum searching}, to appear in Physics Letters A. Also in Lanl-eprint/quant-ph/9906020.
 \end{thebibliography}
 
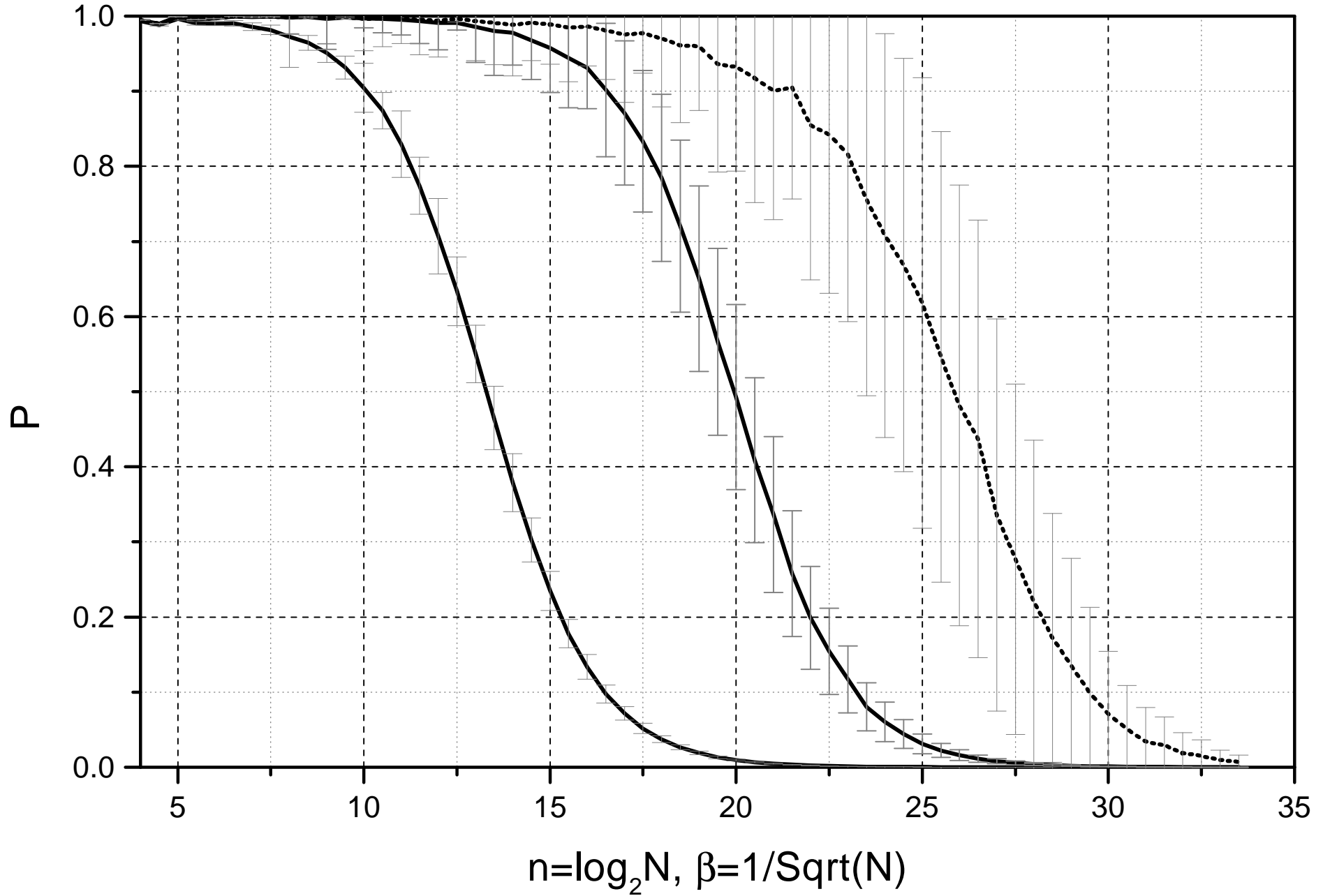
\end{document}